\documentclass[sigconf,screen]{acmart}

\usepackage[bottom]{footmisc}
\usepackage{microtype}

\AtBeginDocument{%
  }

\setcopyright{rightsretained}
\acmDOI{10.1145/3664646.3664759}
\acmYear{2024}
\copyrightyear{2024}
\acmSubmissionID{fsews24aiwaremain-p13-p}
\acmISBN{979-8-4007-0685-1/24/07}
\acmConference[AIware '24]{Proceedings of the 1st ACM International Conference on AI-Powered Software}{July 15--16, 2024}{Porto de Galinhas, Brazil}
\acmBooktitle{Proceedings of the 1st ACM International Conference on AI-Powered Software (AIware '24), July 15--16, 2024, Porto de Galinhas, Brazil}
\received{2024-04-05}
\received[accepted]{2024-05-04}

\newcommand{\FDpipe}{\ensuremath{P}}
\newcommand{\FDboxes}{\ensuremath{\mathcal{B}}}
\newcommand{\FDports}{\ensuremath{\mathcal{P}}}
\newcommand{\FDedges}{\ensuremath{\mathcal{E}}}
\newcommand{\Library}{\ensuremath{\mathcal{L}}}

\newcommand{\FDbox}{b}
\newcommand{\FDport}{p}

\newcommand{\FDboxType}{\ensuremath{boxclass}} %
\newcommand{\FDportType}{\ensuremath{portclass}} %
\newcommand{\FDportBox}{\ensuremath{box}} %
\newcommand{\FDinPoutP}{\ensuremath{src}} %
\newcommand{\FDboxParam}{\ensuremath{param}} %

\newcommand{\type}{\text{type}}
\newcommand{\typeset}{\mathcal{D}}
\newcommand{\functypeset}{\mathcal{F}}
\newcommand{\default}{\text{default}}

\newcommand{\OP}{output port}
\newcommand{\IP}{input port}

\newcommand{\DP}{data port}
\newcommand{\IDP}{input data port}
\newcommand{\ODP}{output data port}

\newcommand{\FP}{function port}
\newcommand{\IFP}{input function port}
\newcommand{\OFP}{output function port}

\begin{document}

\title{Function+Data Flow: A Framework to Specify Machine Learning Pipelines for Digital Twinning}

\author{Eduardo de Conto}
\orcid{0009-0003-9217-0890}
\affiliation{%
  \institution{Nanyang Technological University}
  \city{Singapore}
  \country{Singapore}
}
\affiliation{%
  \institution{CNRS@CREATE}
  \city{Singapore}
  \country{Singapore}
}
\email{eduardo002@e.ntu.edu.sg}

\author{Blaise Genest}
\orcid{0000-0002-5758-1876}
\affiliation{%
  \institution{IPAL}
  \city{Singapore}
  \country{Singapore}
}
\affiliation{%
  \institution{CNRS, CNRS@CREATE}
  \city{Singapore}
  \country{Singapore}
}
\email{blaise.genest@cnrsatcreate.sg}

\author{Arvind Easwaran}
\orcid{0000-0002-9628-3847}
\affiliation{%
  \institution{Nanyang Technological University}
  \city{Singapore}
  \country{Singapore}
}
\email{arvinde@ntu.edu.sg}

\begin{abstract}
    The development of digital twins (DTs) for physical systems increasingly leverages artificial intelligence (AI), particularly for combining data from different sources or for creating computationally efficient, reduced-dimension models. 
    Indeed, even in very different application domains, twinning employs common techniques such as model order reduction and modelization with hybrid data (that is, data sourced from both physics-based models and sensors).
    Despite this apparent generality, current development practices are ad-hoc, making the design of AI pipelines for digital twinning complex and time-consuming.
    Here we propose {\em Function+Data Flow (FDF)}, a domain-specific language (DSL) to describe AI pipelines within DTs. FDF aims to facilitate the design and validation of digital twins.
    Specifically, FDF treats functions as first-class citizens, enabling effective manipulation of models learned with AI.
    We illustrate the benefits of FDF on two concrete use cases from different domains: predicting the plastic strain of a structure and modeling the electromagnetic behavior of a bearing.
\end{abstract}



\maketitle

\section{Introduction}\label{sec:intro}
Digital twins (DTs) are rapidly emerging as a transformative technology for complex systems across diverse industries ~\cite{grievesDigitalTwinMitigating2017}. Examples include applications in smart grids and smart cities~\cite{wangShortTermWindSpeed2023, jafariReviewDigitalTwin2023, danilczykANGELIntelligentDigital2019}, manufacturing~\cite{moyaDigitalTwinsThat2022, ghnatiosHybridTwinBased2024, kritzingerDigitalTwinManufacturing2018a} and aviation~\cite{tuegelReengineeringAircraftStructural2011,utzigAugmentedRealityRemote2019,xiongDigitalTwinApplications2022}. 
The DT market size is projected to grow to US\$ 180 - 250 billion by 2032~\cite{DigitalTwinMarket, EmergingTechnologiesRevenue}.

The overall ambition of DTs is to represent a physical system over its entire lifespan virtually.
To achieve this, twinning leverages simulation and artificial intelligence (AI) for reasoning and decision-making. In addition, a DT can be updated with data to maintain fidelity with the physical counterpart. Despite this ambition, current practices often rely solely on virtual prototypes~\cite{ferriseInteractiveVirtualPrototypes2013} instead of DTs. 
These prototypes are typically created using finite element (FE) modeling, computer-aided design (CAD), or computational fluid dynamics (CFD) frameworks, and they allow predicting (accurately, albeit slowly) the nominal behavior of the system before its physical version is even built. 

As per Grieves~\cite{grievesDigitalTwinMitigating2017}, a virtual prototype can be evolved in three phases of increasing complexity to give rise to DTs.
Firstly, a digital twin prototype ({\em DTP}) is obtained from the virtual prototype using reduced order modeling (ROM) or other techniques. This enables several orders of magnitude faster simulations than the CAD/CFD models~\cite{sancarlosLearningStableReducedorder2021,hartmannModelOrderReduction2018}. 
Secondly, a digital twin instance ({\em DTI}) is created by modeling one particular instance of a physical system.
This uses historical data from sensors placed on that instance, tuning and adapting (offline) the DTP to account for its deviations from the prototype (e.g., manufacturing errors and impact from operating conditions). 
Thirdly, the loop between the physical system and the DTI can be closed by updating (online) the latter using real-time sensor data and controlling the actual instance.
This paper will focus on the first two phases, i.e., the offline design of DTPs and DTIs using ROM and deviation models, respectively. The online exploitation of DTIs also benefits from the methodologies presented in this work (see Section \ref{sec:case-studies}). However, additional advancements specially tailored for real-time update and control are also needed: these advancements will be explored in future research.

Machine learning (ML), a subfield of AI, plays a key role in the design of DTPs and DTIs. 
To obtain DTPs, while established ROM techniques, such as proper generalized decomposition~\cite{chinestaShortReviewModel2011}, can reduce the number of physical variables required to model the system by several orders of magnitude, no physics-based model can work directly on the reduced basis. ML addresses this by enabling the creation of real-time models that operate on the reduced basis. These ML models are trained using simulations generated by the original, slower physics-based model.
On the other hand, developing a DTI requires integrating data from one particular system instance and comparing it with the nominal DTP model. ML plays a crucial role in this process as well. 
Two main approaches are possible. Either (1) ML can be employed to combine the DTP and the specific instance data directly, or (2) ML can be used to create a model of the difference ("ignorance") between the nominal behavior and the actual instance (Hybrid Twins methodology~\cite{chinestaVirtualDigitalHybrid2020}). 
 
We will now describe two {\em motivating examples} that we will use to illustrate the concepts and steps to obtain a DTP or a DTI in two distinct applications.
These are real-world examples based on typical DT applications, as identified in the literature~\cite{chabodDigitalTwinFatigue2022,ghnatiosHybridTwinBased2024}.

\begin{figure}
    \centering
    \includegraphics[width=0.9\columnwidth]{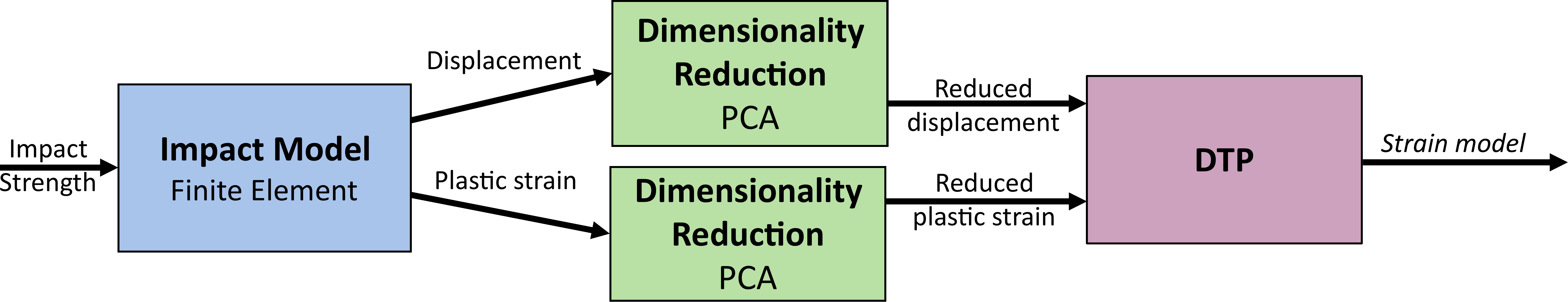}
    \caption{Pipeline for structural health monitoring.}
    \label{fig:pipe-strain-overview}
    \Description{A pipeline for structural health monitoring. Four boxes are shown. The impact model is connected to the two PCA boxes and the output of each PCA is connected to the DTP.}
\end{figure}

{\em Structural Integrity Monitoring.} 
The first use-case focuses on structural health monitoring~\cite{chabodDigitalTwinFatigue2022}, enabling, e.g., predictive maintenance~\cite{tuegelReengineeringAircraftStructural2011}.
Here, the goal is to predict the plastic strain of a certain structure given an observed deformation. 
While currently no (non-destructive) methodology can directly measure the plastic strain, 3D images of the observed deformation can be collected via digital image correlation~\cite{johnsonComplexitiesCapturingLarge2023,tehraniPipeProfilingUsing2020}.
To monitor the structural strains, the following pipeline, illustrated in Figure~\ref{fig:pipe-strain-overview}, could be used: 

\begin{enumerate}
    \item Use a (slow) finite element (FE) impact model to simulate different impact strengths 
    and obtain deformation and plastic strain values as output.
    \item Reduce the deformations and plastic strains using principal component analysis (PCA).
    \item Train a DTP (using supervised learning and the above-reduced dataset) to learn a model that predicts the reduced plastic strain from the reduced deformation.
\end{enumerate}

\begin{figure}
    \centering
    \includegraphics[width=0.9\columnwidth]{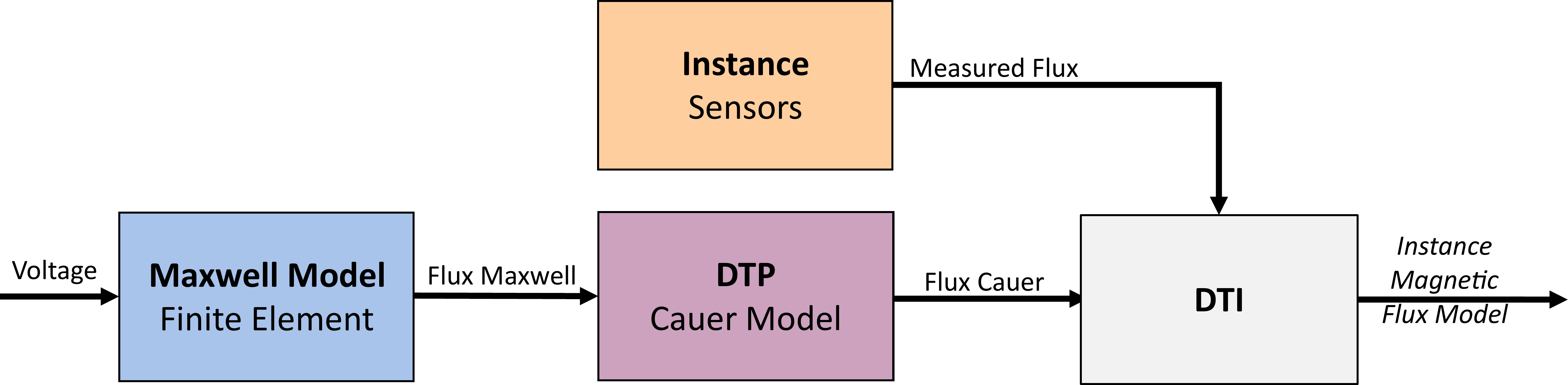}
    \caption{Pipeline to model an active magnetic bearing.}
    \label{fig:magnetic-bearing-overview}
    \Description{A pipeline to model an active magnetic bearing.
    Four boxes are shown. The Maxwell model is connected to the DTP. Next, a DTI receives data from the DTP and sensor data.}
\end{figure}

{\em Electromagnetic Bearing Modeling.} 
The second use-case relates to the modeling of an active magnetic bearing~\cite{sivasrinivasApplicationActiveMagnetic2018},
facilitating real-time analysis of the device's behavior.
The goal is to predict the induced magnetic flux based on the voltage applied to the device. 
A recent work ~\cite{ghnatiosHybridTwinBased2024} proposed to combine a slow FE model with a ROM (Cauer) to achieve a fast and accurate pipeline. 
The following pipeline, depicted in Figure~\ref{fig:magnetic-bearing-overview}, is used:

\begin{enumerate}
    \item Leverage an FE model based on the Maxwell Equations to accurately calculate the magnetic flux ({\em Flux Maxwell}) within the system based on the applied voltage. This model is computationally expensive.
    \item Obtain a faster DTP (denoted as {\em Cauer Model}) allowing for much faster simulations. 
    This DTP is a model of the magnetic bearing, described as an electric circuit with several branches of resistor and inductance elements. The output of this step is {\em Flux Cauer}, the magnetic flux induced within the equivalent circuit, a (linear) approximation of Flux Maxwell.  
    \item Integrate historical data from a specific magnetic bearing instance (e.g., sensor measurements) %
    to obtain the DTI.
\end{enumerate}

\subsubsection*{Challenges in Specifying ML Pipelines for Twinning}
In conventional data-driven pipelines used in classical ML applications, the objective is to train {\em one} model to accomplish {\em one} task. 
The model does not need to be manipulated, and many operations can be implicit (e.g. the training and the inference of models/functions do not need to be distinguished).
Given that the application of ML to DTP/DTI design is a relatively new research area, the development practices, methodologies, and tools, are not yet fully established. Compared to the standard ML pipelines, the following challenges arise:

{\em Distinction Between Training and Inference.}
In supervised learning, training the model and running inference from the model are different steps of the process. For instance, while the latter (inference) takes input data $X$ (e.g. voltage) and outputs data $Y$ (e.g. predicted flux), the former (training) receives pairs $(X, Y)$, where $Y$ is the ground truth answer from the query $X$. Note that Figure~\ref{fig:magnetic-bearing-overview} lacks a clear distinction between {\em learning} the Cauer Model and {\em inferring the flux} predicted by the Cauer. Also, the input of each step is implicit. 

{\em Function/Model Manipulation.} 
By contrast with standard ML pipelines, twinning requires several models/functions. These functions include projection to a reduced basis, several DTPs, and the DTI itself, where some functions are instrumental in generating others. Hence, operations on these models must be either explicit or they must adhere to a very rigid implicit pipeline. 

{\em Pipeline Diversity.} 
Although twinning involves the same basic operations (reduced order modeling and combining data from different sources), the specific pipelines employed heavily depend on the application domain or even the individual application. Figures~\ref{fig:pipe-strain-overview} and~\ref{fig:magnetic-bearing-overview} illustrate this variability. 
In the electromagnetic bearing pipeline (Figure~\ref{fig:magnetic-bearing-overview}), both the DTP (Cauer Model) and the FE (Maxwell Model) have the same input/output.
However, the structural health monitoring pipeline (Figure~\ref{fig:pipe-strain-overview}) demonstrates a situation in which the {\em input} of the DTP (displacement) is an {\em output} of the FE impact model. That is, the pipeline is not directly reducing the Impact Model.

\subsubsection*{Proposal.} 
To address the challenge of {\em pipeline diversity}, we adopt the visual dataflow paradigm~\cite{johnstonAdvancesDataflowProgramming2004}, allowing for an intuitive and adaptable pipeline description. %

Concerning {\em function/model manipulation}, we propose a novel Function+Dataflow (FDF). 
FDF is a domain-specific language (DSL) for the specification of ML pipelines used for DT design. 
FDF enables the manipulation of functions learned by the ML pipeline. To achieve this, FDF extends the traditional dataflow by incorporating {\em functions as first-class citizens}.
The function is defined as another type of flow besides data streams, as in \cite{fukunagaFunctionsObjectsData1993}, a generic data-flow programming language, as well as in dataflow-based DSLs for different contexts (specifically, quantum-classical combination~\cite{sivarajahTierkreisDataflowFramework2022} and data engineering~\cite{demeoDomainSpecificVisualLanguage2022}). 
This addition allows us to:
\begin{itemize}
    \item {\em Decouple} the learning of a function (e.g. how to project data into a PCA reduced basis) and the usage of the function (actual projection of data into the reduced basis).
    \item {\em Manipulate} the models explicitly, allowing their use and reuse as necessary. For example, in the structural health monitoring case, we can reuse the reduced basis projections. 
    \item {\em Infer and track} the input and output data type for each function within the pipeline automatically. This capability allows for implicit type-checking and can offer significant benefits to the users: we can suggest valid inputs to the users, or warn them of potential incompatibilities, avoiding bugs. Notice that actual data types are never required from the user, avoiding a tedious process.
\end{itemize}

\section{Related Works}\label{sec:related}

\subsubsection*{Machine Learning Workflow}\label{sec:ai-workflow}

Recent advancements and the deployment of AI and ML in critical applications have led to a surge of tools to structure ML pipelines (e.g., Kedro~\cite{alamKedro2024}, MLflow~\cite{chenDevelopmentsMLflowSystem2020}, Apache Airflow\cite{ApacheAirflow}, Kubeflow~\cite{Kubeflow}). These tools allow the description of a machine learning pipeline and, in addition, allow, among other things, tracking experiment results, performing model versioning, and monitoring the model performance. The ML pipeline in these systems is typically represented by directed acyclic graphs.

Despite their advantages, the current tools still require a significant integration effort: they focus on generating a single ML model, which is not explicitly represented in the pipeline~\cite{lwakatareDataScienceDriven2020}. One exception is the Transformers Library~\cite{wolfHuggingFaceTransformersStateoftheart2020}, but this library only supports predefined pipelines for common tasks in the natural language processing domain (object detection, summarization, etc.). In all cases, the essential task of the workflow is to produce \textit{one} model which can only be recovered and used externally after the learning is completed.

As described in the "Function/Model Manipulation" challenge, the design of DTs involves the creation and manipulation of multiple interrelated models and functions. 
For instance, the DTP may be needed to create the DTI, etc. 
A possible way to handle model manipulation within dataflow is to transmit the model as a data token. 
While technically feasible, this method discards valuable information (e.g. the input/output type of the models).

To overcome these limitations, FDF includes a {\em dedicated function flow} (used to transmit learned models and functions). Valuable information about the functions learned in the ML workflow can thus be preserved and transmitted. This includes the data types accepted as input and produced as output.

\subsubsection*{Twin Builders from CAD/CFD tools}\label{sec:cad}

Commercial software vendors specializing in CFD/CAD, such as Ansys and Siemens, have incorporated ROM capabilities into their existing software toolkits (e.g., Ansys Twin Builder~\cite{AnsysTwinBuilder}, Siemens Simcenter Amesim~\cite{SimcenterSystemsSimulation}). These functionalities typically follow a similar workflow: execute simulations within a CAD/CFD environment and export them to a dedicated ROM module with limited user-controllable parameters. 
Notably, none of these tools offer an open, dataflow-based pipeline that can be customized for specific use cases.
Therefore, the results are often inconsistent and heavily dependent on whether the predefined and proprietary workflows available are suitable for the domain of interest. 
The core novelty of our methodology is its ability to make the ROM pipeline {\em fully customizable} using FDF. 
This flexibility is crucial for adapting the model to diverse application contexts. Importantly, our approach can still offer predefined, yet fully customizable workflow templates.

\section{Function+Data Flow Syntax}\label{sec:pipeline-syntax}
We now describe the syntax of the Function+Data Flow (FDF) pipeline. We first provide an overview of the rationale and visual syntax of the components of the pipeline, and then formalize it.

\subsection{Syntax Overview}
\begin{figure}[t!]
    \centering
    \includegraphics[width=0.6\columnwidth]{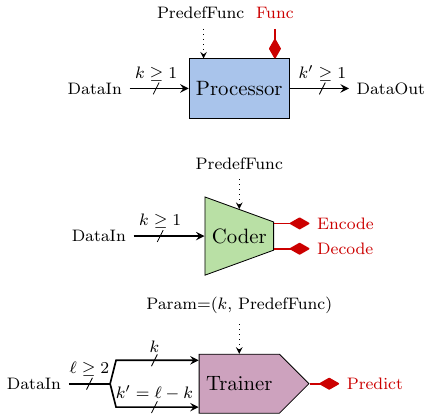}
    \caption{Visual syntax for boxes of Function+Data Flow: Processor on top, Coder in the middle, and Trainer at the bottom. The processor executes either a function $Func$ learned by an earlier box in the pipeline or a {\em predefined} function $PredefFunc$. The value $k$ in the Trainer's Param specifies the number of input ports to consider as $X$. The remaining ports are the $Y$ in the $(X, Y)$ supervised learning pairs.}
    \label{fig:box-syntax}
    \Description{Refer to the main text for a detailed description.}
\end{figure}

An FDF pipeline is composed of boxes (as shown in Figure~\ref{fig:box-syntax}) that represent the different processing steps. 
Each box has several input and output ports. There are two types of ports: function ports (in {\color{red}red}), sending/receiving a single {\em learned function}, and data ports (in black) sending/receiving a batch of data. 

There are three user-specified boxes in FDF, each associated with a different task in the twinning workflow:
\begin{itemize}
    \item Processor: for typical data processing (including applying functions learned by other boxes),
    \item Coder: for learning a reduced basis (or unsupervised clustering) and the associated projection and inverse projection, and
    \item Trainer: for learning a function with supervised ML.
\end{itemize}

\begin{table*}[ht!]
    \centering
    \caption{Summary of Box Syntax}
    \label{tab:box-syntax}
    \small
\begin{tabular}{llllll}
    \toprule
    & \multicolumn{5}{c}{Box Type}                                                   \\ 
    \cmidrule{2-6}
    Characteristics & Processor & Coder & Trainer & FuncOut & DataIO \\ \midrule
    Representation  & \parbox[t]{2cm}{Light blue \\ rectangle} 
                    & \parbox[t]{2cm}{Pale green \\ trapezoid}       
                    & \parbox[t]{2cm}{Pale violet \\ pentagonal} 
                    & Invisible 
                    & Invisible \\
\midrule
    Application     & \parbox[t]{2cm}{Data \\ processing}      
                    & \parbox[t]{2cm}{Unsupervised \\ learning} 
                    & \parbox[t]{2cm}{Supervised \\ learning}   
                    & \parbox[t]{2cm}{Function \\ source/sink} 
                    & \parbox[t]{2cm}{Data \\ source/sink} \\
\midrule
    Inputs Data Ports    & One or more  & One or more & Two or more & Zero & Zero or more   \\
    Inputs Function Ports & Zero or one  & Zero        & Zero        & Zero or more & Zero  \\
\midrule
    Output Data Ports    & One or more & Zero & Zero & Zero & Zero or more    \\ 
    Output Function Ports    & Zero & One or Two & One & Zero  & Zero    \\ 
\bottomrule
\end{tabular}

\end{table*}

Each box is represented by different polygon shapes, as shown in Figure~\ref{fig:box-syntax}. There are also two implicit (i.e., non-depicted) boxes, namely FuncOut and DataIO, that represent the pipeline's external input/output in terms of functions and data, respectively. 

We now outline the syntax of each pipeline box. A summary is provided in Table~\ref{tab:box-syntax}, their visual representation is given in Figure~\ref{fig:box-syntax} and a description is given in the following.

The {\em Processor boxes} are represented by a light blue rectangle, as in the top of Figure~\ref{fig:box-syntax}. It has multiple \IDP s to receive the data for processing, 
and multiple \ODP s to return the processed data. 
The function to execute is either provided through an \IFP~(function learned by the previous boxes) or is a predefined function from a library, as specified by box parameter. 

The {\em Coder boxes} are represented by a pale green trapezoid, as in the middle of Figure~\ref{fig:box-syntax}. It has multiple \IDP s to receive the data from which to compute the reduced basis and one or two \OFP s to return the encoder and decoder functions (i.e., the projection and inverse projection onto the reduced basis, respectively). 
Coder boxes have no \IFP: the specific encoding/decoding algorithm is a predefined function from a library, as specified by the parameter of the box. For instance, "PCA (99\%)" specifies a principal component analysis capturing $\geq 99\%$ of the variance of the original data.

The {\em Trainer boxes} are represented by a pale violet pentagon, as in the bottom of Figure~\ref{fig:box-syntax}. 
It has $\ell \geq 2$ \IDP s to receive the supervised $(X, Y)$ pairs, with $k < \ell$ ports providing $X$ and the remaining $k'=\ell-k$ providing $Y$, and it has one \OFP~ to return a function that predicts $Y$ given $X$. The number $k$ is provided in the first component of the Trainer box parameter. %
The Trainer boxes have no \IFP: %
the specific ML training algorithm is provided in the second component of the Trainer box parameter and is a predefined function from a library (e.g. PyTorch). For instance, "NN (50, 50, SGD)" indicates that stochastic gradient descent (SGD) shall be used to learn a feedforward neural network with 2 hidden layers, each with 50 nodes. 

Finally, the \textit{implicit boxes (FuncOut and DataIO)} represent %
the input/output dependencies of the pipeline. 
{\em FuncOut} has no input/output data ports, no \OFP, but it can accept any number of \IFP s. A function is sent to FuncOut to export it to external pipelines.
{\em DataIO} has no input/output function port, and it has any number of data input/output ports. The output ports are sources for the different data batches used by the FDF pipeline. The input ports are data sinks, i.e., they store data, allowing them to be persisted in disk for further analysis. 

\subsection{Formal Syntax}\label{sec:formal-syntax}
An FDF pipeline is defined as $\FDpipe=(\FDboxes', \FDboxType, \FDports,\FDportType, \FDportBox,\allowbreak \FDinPoutP, \FDboxParam)$, where:

\begin{itemize}
    \item $\FDboxes =  \{\FDbox_1, \ldots , \FDbox_{n}\}$ are the user-defined boxes.
    We denote $\FDboxes' = \FDboxes \sqcup \{DataIO,FuncOut\}$.
    \item $\FDboxType: \FDboxes \rightarrow \{\text{Processor, Coder, Trainer}\}$ defines the class of each box.
    \item $\FDboxParam: \FDboxes \rightarrow \Library \, \cup \,  \mathbb{N} \times \Library$ provides the parameters of a given box, where $\Library$
    is a library of predefined functions.
    \item $\FDports = \FDports^I \sqcup \FDports^O = \{1, \cdots, m\}$ is the ordered set of natural numbers up to $m$, the number of ports. It is partitioned into the sets of input ports $\FDports^I$ and output ports $\FDports^O$.
    \item $\FDportType: \FDports \rightarrow \{\text{Data, Function}\}$ 
    provides the class (Data or Function) of each port.
    \item $\FDportBox: \FDports \rightarrow \FDboxes'$ is a function associating each port with the box it belongs to.
    \item $\FDinPoutP: \FDports^I \rightarrow \FDports^O$ is a function associating every input port with the output port providing its data/function. 
    $\FDinPoutP$ is such that the class of an input and the associated output ports are the same: 
    $\forall \FDport \in \FDports^I, \FDportType(\FDport) = \FDportType(\FDinPoutP(\FDport))$.
\end{itemize}

\subsection{Example}

\begin{figure}[h]
    \centering
    \includegraphics[width=0.6\columnwidth]{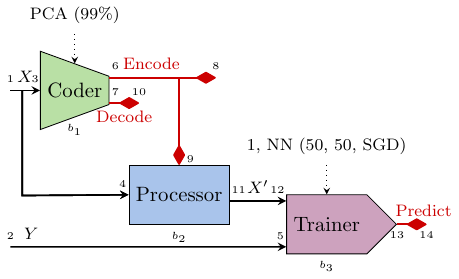}
    \caption{Minimal FDF pipeline with annotations}
    \label{fig:pipe_annotated}
    \Description{Detailed description given in text.}
\end{figure}

We now review a minimal FDF pipeline to show the one-to-one correspondence between the visual syntax (Figure~\ref{fig:box-syntax}) and the formal syntax just described. The pipeline is given in Figure~\ref{fig:pipe_annotated} and is described in the following:
\begin{enumerate}
    \item Use a Coder box with a predefined function "PCA (99\%)" to obtain a PCA basis of $X$ with $99\%$ of accuracy,
    \item Use a Processor box executing the learned function "Encode" and retrieve the reduced data $X'$ from (full) $X$,
    \item Use a Trainer box with a Param "1, NN (50, 50, SGD)" to learn a function $Predict$ that predicts $Y$ from $X$. This function is learned with stochastic gradient descent on a neural network with 2 hidden layers of 50 nodes each. 
\end{enumerate}

Here is the formal definition of this FDF pipeline $\FDpipe$:
\begin{itemize}
    \item $\FDboxes = \{\FDbox_1, \FDbox_2, \FDbox_3\}.$
    \item $\FDboxType(\FDbox_1) =$ Coder; 
          $\FDboxType(\FDbox_2) =$ Processor; \\
          $\FDboxType(\FDbox_3) =$ Trainer.
    \item $\FDboxParam(\FDbox_1) =$ PCA($99\%$); 
          $\FDboxParam(\FDbox_3) =$ 1, NN (50, 50, SGD)
    \item $\FDports = \FDports^I \sqcup \FDports^O$, with 
          $\FDports^I = \{3, 4, 5, 8, 9, 10, 12, 14\}$, 
          $\FDports^O = \{1, 2, 6, 7, 11, 13\}$.
    \item $\FDportType(p) = \begin{cases}
        (Data), & \text{if } p \in \{1, 2, 3, 4, 5, 11, 12\} \\
        (Function), & \text{if } p \in \{6, 7, 8, 9, 10, 13, 14\}\\
    \end{cases}$      
    \item $\FDportBox(p) = \begin{cases}
        DataIO,   & \text{if } p \in \{1, 2\} \\
        FuncOut,   & \text{if } p \in \{8, 10, 14\} \\
        \FDbox_1, & \text{if } p \in \{3, 6, 7\} \\
        \FDbox_2, & \text{if } p \in \{4, 9, 11\} \\
        \FDbox_3, & \text{if } p \in \{5, 12, 13\}\\
    \end{cases}$
    \item $\FDinPoutP(3) = 1;
           \FDinPoutP(4) = 1;
           \FDinPoutP(5) = 2;
           \FDinPoutP(8) = 6;
           \FDinPoutP(9) = 6;
           \FDinPoutP(10) = 7;
           \FDinPoutP(12) = 11;
           \FDinPoutP(14) = 13.$
\end{itemize}

\section{Function+Data Flow Semantics}\label{sec:pipeline-semantics}
To define the semantics of an FDF pipeline, we first introduce the directed FDF graph associated with an FDF pipeline, which will define an explicit order in which the FDF pipeline is executed.

\subsection{FDF Graph} 

The FDF directed graph $G(\FDpipe) = (\FDports, \FDedges)$
associated with an FDF pipeline $\FDpipe = (\FDboxes', \FDboxType, \FDports, \FDportType, \FDportBox, \FDinPoutP, \FDboxParam)$ is defined as follows. 
For all ports $p,q \in \FDports$, we have $(p,q) \in \FDedges$ iff either:

\begin{itemize}
    \item $p \in \FDports^O, q \in \FDports^I$ and $p= \FDinPoutP(q)$. These are the edges between the boxes, or
    \item $p \in \FDports^I, q \in \FDports^O$ and $\FDportBox(p)=\FDportBox(q) \in \FDboxes$, which excludes the two implicit boxes $DataIO, FuncOut$. This models the {\em (complete)} internal dependencies in each box.
\end{itemize}

Note that output ports associated with the implicit box $DataIO$
will have no predecessor in $G(\FDpipe)$, and input ports associated with the implicit boxes $DataIO, FuncOut$ will have no successors.

We call the directed graph $G(\FDpipe)$ and the FDF pipeline $\FDpipe$ {\em well-formed} if $G(\FDpipe)$ is a directed acyclic graph (DAG). To be executable, $\FDpipe$ needs to be well-formed. We will thus assume in the following that $\FDpipe$ is well-formed. Note that this can be tested in linear time.

\emph{Example.} 
In Figure~\ref{fig:pipe_annotated}, we have the following edges:
\begin{align*}
    \FDedges = \{
      & (1, 3), (1, 4), (2, 5), (6, 8), (6, 9), (7, 10), (11, 12), (13, 14), \\ 
      & (3, 6), (3, 7), (4, 11), (9, 11), (5, 13), (12, 13), (13, 14) \}
\end{align*}

\subsection{FDF Execution}

Well-formedness allows us to decide the order in which to execute boxes, which boxes may be executed in parallel, and which must be executed before/after another.
We say that $b \in \FDboxes$ (so not an implicit box) is a direct predecessor of $b'$, 
denoted $b \lessdot b'$ whenever $\exists (p,q)$ with $p= \FDinPoutP(q)$, $\FDportBox(p)=b$ and $\FDportBox(q)=b'$. The execution of the box $b'$ is blocked until all the explicit boxes $b\lessdot b'$ have been executed. 
In addition, a box executes only if the batches for all its \IDP s have the same number of samples. 
The execution semantics depend on the class $\FDboxType(\FDbox)$ of the FDF box $\FDbox$. We describe in the following how each box is executed in FDF.

\subsubsection*{Processor}

Running a Processor box (i.e., $\FDboxType(b)$ = Processor) with $k$ input data ports and $k'$ output data ports consists of the following steps: 

\begin{enumerate}
    \item Load the function $f$ to execute and the input data DataIn. 
    The function $f$ is provided either as a predefined function
    PredefFunc, defined by the parameter, or as an input function port (in which case $f$ is a learned function, the output of a previous Coder or Trainer box of the pipeline). DataIn are batches in the \IDP s~of $b$.
    \item Execute $f$ for each sample of the batch DataIn (granted that $f$ accepts a vector of size $k$ as input and produces a vector of size $k'$ as output).
    \item Return the processed DataOut in the $k'$ \ODP s of the Processor. 
\end{enumerate}

\subsubsection*{Coder} 
Running a Coder box (i.e., $\FDboxType(b)$ = Coder) with two output function ports consists of the following steps:
\begin{enumerate}
    \item Load the DataIn from the \IDP. 
    \item Load the predefined function PredefFunc, provided by the parameter of $b$, specifying how to obtain the reduced basis from a batch of data.
    \item Run PredefFunc on DataIn to obtain the 2 functions Encode and Decode\footnote{Some algorithms specified in PredefFunc may only provide one of these two functions. Thus, Coder may have only one function output in some cases.} 
    based on the reduced basis. Return these functions in their corresponding \OFP.
\end{enumerate}

\subsubsection*{Trainer}

Running a Trainer box (i.e., $\FDboxType(b)$ = Trainer) with $\ell$ \IDP s consists of the following steps:
\begin{enumerate}
    \item  Load $Param=(k, PredefFunc)$ defining a number $k < \ell$ and a predefined function PredefFunc. The number $k$ is used in the next step to distinguish $X$ values from $Y$ values in the $(X,Y)$ supervised learning pairs. 
    PredefFunc specifies how to learn the model %
    from the supervised pairs $(X, Y)$.
    \item Load the required batch of $(X,Y)$ pairs, with $X$ obtained from the first $k$ \IDP s, and $Y$ from the last $k'=\ell-k$ \IDP s.
    \item Run PredefFunc on the batch of $(X,Y)$ pairs to obtain the function Predict for the generalization of $X \mapsto Y$.
\end{enumerate}

\section{Implicit Typing}\label{sec:implicit_typing}
In FDF, we can leverage the fact that functions are first-class citizens to infer extra information about them. Here we explain how to automatically infer (implicit) types for the data and functions generated within the pipeline, based on the FDF syntax and semantics.

Implicit typing endows the FDF pipeline with advantages commonly associated with statically typed languages, e.g. being less prone to errors and easier to maintain~\cite{bognerTypeNotType2022,rayLargeScaleStudy2014}. 
The typing is {\em implicit}, that is, it does not require the user to manually define explicit types, which can be laborious~\cite{oreAssessingTypeAnnotation2018}, and
even infeasible in the case of DT design due to the dynamic nature of the functions learned within the pipeline. For instance, when applying PCA to obtain a reduced basis preserving 99\% of the variance of the original dataset, the explicit output type (i.e., number of dimensions) depends on the training data: explicit typing would not be feasible in this case. 

\subsection{Implicit Types}

First, for all (data or function) port $i \in \FDports = \{1, \ldots, m\}$, 
we define $\default(i)=i$. Now, we associate an {\em implicit} data type $\type(p)$ to each \DP~and to each \FP. 

For {\em \DP s}, we define the set 
$\typeset = \{1, \ldots, m\}$ of Data Types, which is the same as the set of ports 
$\FDports$. 
By default, port $p$ has implicit type $\type(p) \gets \default(p)$, but it may be given type 
$\type(p) \gets \default(q)< \default(p)$ if it is known that the types of ports
$p,q$ are the same. In general, some numbers in $[1, m]$ will not be used, as several ports will have the same implicit type.

For {\em \FP s}, the set of function types is defined as 
$\functypeset = \cup_{i \in \mathbb{N}} \typeset^i \times \cup_{j \in \mathbb{N}} \typeset^j$.
That is, a function $f$ with $k$ inputs and $k'$ outputs will have the implicit type 
$((\type_{1}, \ldots, \type_{k}), (\type_{k+1},$ \newline$\ldots , \type_{k+k'}))$, 
where $\type_{i \leq k}$ is the implicit type of the $i$-th input,
and  $\type_{k+i}$ is the implicit type of the $i$-th output of $f$ for $i\leq k'$.

By default, a {\em Coder} box $\FDbox$, with 
$\type_{1}, \ldots, \type_{k}$ denoting the types of \IDP s of $\FDbox$, generates:
\begin{itemize}
    \item a function Encode of type
    $((\type_{1}, \ldots, \type_{k}),(\type_{Out}))$. $\type_{Out}$ is a fresh type never seen before and represents the data on the reduced basis, and
    \item a function Decode of type $((\type_{Out}),(\type_{1}, \ldots, \type_{k}))$.
\end{itemize}
This default can be changed by providing extra information in the library containing the predefined function specified by the Coder's parameter. For example, if the parameter of $\FDbox$ calls a normalization procedure, we could have $\type_{Out} = \type_{1}$.

A {\em Trainer } box $\FDbox$ with $\ell$ \IDP s, generates, by default, a function Predict of type
$((\type_1, \ldots, \type_k),(\type_{k+1}, \ldots,\type_{\ell})),$
where $\type_{1}, \ldots, \type_{\ell}$ are the types of the \IDP s of $\FDbox$, and $k$ is the number provided in the first component of the parameter of the Trainer box $\FDbox$.
This default can also be changed in the library.

\subsection{Type propagation and checking}

We now explain how to propagate the types automatically between ports. 
The types are propagated via the FDF Graph topological order. We assume, without loss of generality, that the port numbering follows the topological order. 

Note that the type checking may return warnings to the user if it does not have 
enough information to ensure that the two types are equal.
In this case, the user would either confirm that the two ports have the same type or rectify the pipeline if the type mismatch is genuine. 
The user can also add explicit {\em type annotations} before running the type checking, to provide this information. Specifically, if different ports $\FDport_1, p_2, \ldots, p_s$ have the same annotation (except exponents), then they have the same implicit type. Thus, we can set $\type(p_{1}) = \ldots = \type(p_{s}) \gets min_{j \leq s}(\type({p_j}))$.

The type propagation proceeds in three main steps. The {\em first step} is to propagate types for the DataIO \ODP. Let $\{1\ldots, r\}$ be the \ODP s of DataIO (that is, the first $r$ ports of the FDF pipeline). By default, 
each port $i \leq r$ will have a different data type: 
$\type(i) \gets \default(i) = i$. The user may add annotations to specify otherwise.

The {\em second step} is to propagate the types through the ports which are associated with a box $b$. First, each \IP 
~$p \in \FDports^I$ with $\FDportBox(p)=b$ copies the implicit data type from the corresponding \OP~$\FDinPoutP(\FDport)$, that is $\type(\FDport)\gets \type(\FDinPoutP(\FDport))$. 

Finally, the {\em third step} is to compute the implicit type for each \OP~of $\FDbox$. Note that, in general, an \OP~$p$ can either have its default type, $\type(p) = \default(p)$, or it can have a type $\type(p) = \type(q) < \default(p)$, propagated from a previous port $q$ (through possibly several boxes). The implicit type depends on $\FDboxType(\FDbox)$ as follows. 

\subsubsection*{Coder Type Propagation}

Let $\FDport_{Out}$ be the \OFP~ of $\FDbox$ (if there are two output function ports, take the minimal $\default(\FDport)$: there is at least one \OFP). Then, a box $\FDbox$ with $\FDboxType(\FDbox)=$~Coder has $\default(\FDport_{Out})$ as the output type of Encode (and the input type of Decode). This guarantees by construction that this type has not been used before.
Let $\type_1, \ldots, \type_r$ be the implicit types of the \IDP s of $\FDbox$, and let $p_{\text{Encode}}$, $p_{\text{Decode}}$ be the two output function ports. We define: 
\begin{align*}
    \type(p_{\text{Encode}}) &\gets ((\type_1, \ldots, \type_r),(\default(\FDport_{Out}))) \\
    \type(p_{\text{Decode}}) &\gets ((\default(\FDport_{Out})),(\type_1, \ldots, \type_r)) 
\end{align*}

\subsubsection*{Trainer Type Propagation}
A box $\FDbox$ with $\FDboxType(\FDbox)=$~Trainer has a single \OFP. 
We define its implicit type as follows. Let $k$ be the number provided by Param; let $\type_1, \ldots, \type_\ell$ be the implicit types of the \IDP s of $\FDbox$; let $\FDport$ be the \OFP~of $\FDbox$. Then:
$$\type(\FDport) \gets ((\type_1, \ldots, \type_k),(\type_{k+1}, \ldots, \type_\ell))$$

\subsubsection*{Processor Type Propagation}

For $\FDboxType(\FDbox)=$~Processor, we have two cases. 
The {\em first case} is when $\FDbox$ has a \IFP~$\FDport_F$. We denote $\type(\FDport_F)= ((\type_1, \ldots,\type_k),\allowbreak(\type'_{1}, \ldots, \type'_{k'}))$. Before propagating the type, we must ensure the following conditions are absent. If a condition is detected, we raise an error or warning:

\begin{enumerate}
    \item {\em Mismatch in the number of input/output.} A mismatch error occurs if the number of \IDP s of $\FDbox$ is not $k$ or if the number of \ODP s of $\FDbox$ is not $k'$. The user should fix the pipeline.
    \item {\em Inconsistent input type.} 
An inconsistent input type warning occurs when 
$\exists j \in [1, k]: \type(\FDport_j) \neq \type_j$, i.e., 
the type expected as input by $\FDport_F$ does not match the type of port $p_j$. %
To fix this inconsistency, the user can either tell that the two types are identical (for instance, by annotation) or fix the pipeline.
\end{enumerate}

If no such problem is encountered, we can set $\type(\FDport_{k+j}) \gets \type'_{j}$ for all $j \leq k'$, and propagate to the next boxes.

\medskip 

The {\em second case} is when $\FDbox$ has no \IFP, but instead, its parameter specifies a predefined function PredefFunc from a library. The library is unaware of the implicit types propagated in one particular FDF pipeline. Further, 
a function from a library can be polymorphic, accepting several types as input, another reason to not impose strong typing.
The library can provide however weak type information:
first, its number $k$ of input ports and $k'$
of output ports (If the input vectors can take any size, then the input is multiplexed into a single port and $k=1$; similarly for the output). The library can also specify a partition $P_1, \ldots, P_r$ with $P_1 \sqcup P_2 \sqcup \cdots \sqcup P_r$ of the set of all (input and output) ports of the function. Each partition represents the fact that types should be equal within the partition. 

Similar to the first case, we must first ensure the following two conditions are absent:
\begin{enumerate}
    \item Mismatch in the number of inputs or outputs.
    \item Inconsistent input type. This condition occurs if two ports from the same partitions have different types, i.e., if 
    $\exists i \leq r: (p, q) \in P_i\text{ and }\type(p)\allowbreak\neq\type(q)$.
\end{enumerate}

If a warning is raised, the user can fix it as in the first case, e.g. by providing the information that the two types are the same. 
Once there are no more warnings or errors, the type propagation proceeds as follows. For all $i \leq r$, either:
\begin{enumerate} 
    \item there is an \IDP~$p \in P_i$: we set $\type(q)\gets\type(p)$
    for all the output ports $q$ in $P_i$, as the library specified that these output data ports have the same implicit type as the input data port $p$, or %
    \item there is no such \IDP: we set a fresh $\type(q)\gets\min_{q \in P_i}\default(q)$  for all the output ports $q$ in $P_i$.
\end{enumerate}

\section{Application to Motivating Examples}\label{sec:case-studies}
Now we illustrate how the FDF formalism can be applied to the motivating examples described in Section~\ref{sec:intro}.

\subsection{DTP for Material Strain Prediction}

\begin{figure}[b]
    \centering
    \includegraphics[width=\columnwidth]{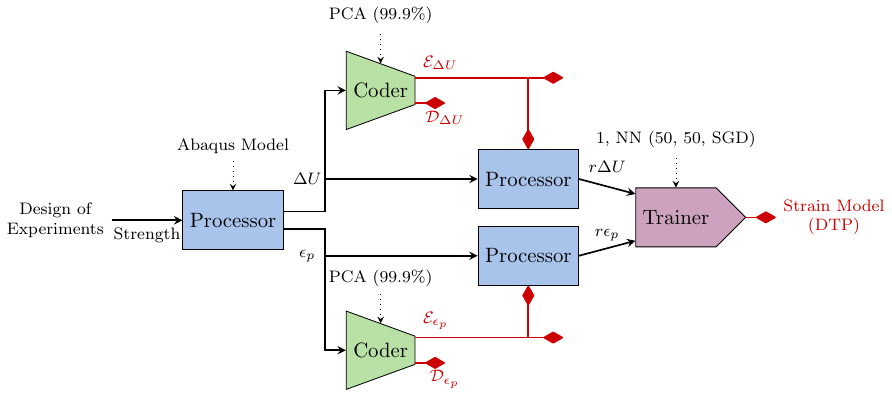}
    \caption{An FDF pipeline to learn a Strain Model DTP.}
    \label{fig:pipe_cetim_offline}
    \Description{An FDF pipeline to learn a DTP that predicts the plastic strain of a material given an observed deformation.}
\end{figure}

Recall that the first motivating example (see Figure~\ref{fig:pipe-strain-overview}) aims to predict the plastic strain of a structure from an observed deformation~\cite{chabodDigitalTwinFatigue2022}.  
Figure~\ref{fig:pipe_cetim_offline} describes an FDF pipeline generating a Strain Model DTP. 

From a design of experiments exploring impacts with different strengths and repetitions, we obtain a correlation between the deformation $\Delta U$ and the plastic strain $\epsilon_p$ using a (slow) finite element model (first Processor box).

Both these datasets (set of deformations and set of strains) have high dimensionality (>1000 dimensions). 
First, we reduce the dimensions of each dataset, using PCA to learn a reduced basis (tens of dimensions) allowing us to recover $99.9\%$ of the precision. The learning of the PCA basis (in the Coder box, that is generating the encoding $\mathcal{E}$ and decoding $\mathcal{D}$ functions) is decoupled from applying it in the second Processor box. 
This produces a reduced dataset $r\Delta U$.
The same is true for the reduced plastic strain $r\epsilon_p$. 
Notice that the $i$-th reduced plastic strain corresponds to the $i$-th reduced deformation. 
Last, we learn a neural network (with 2 layers of 50 nodes each) generalizing the function from reduced displacement $r\Delta U$ to reduced strain $r\epsilon_p$.

\subsubsection*{Exploitation}

Once the different functions have been learned, the DTP can be exploited to obtain the plastic strain from actual 3D images of a deformation. An exploitation pipeline can be described in FDF as well: the pipeline is composed of a series of Processor boxes using the learned functions. 

\begin{figure}[t]
    \centering
    \includegraphics[width=\columnwidth]{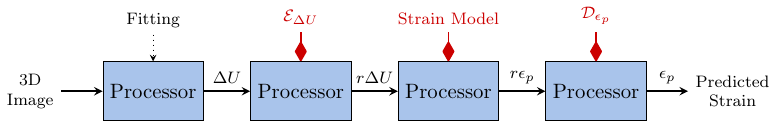}
    \caption{Exploitation pipeline for Material Strain Prediction.}
    \label{fig:pipe_cetim_online}
    \Description{Four processor boxes are shown. Detailed description in text.}
\end{figure}

The pipeline, illustrated in Figure~\ref{fig:pipe_cetim_online}, starts by fitting the 3D image on the finite element mesh of the structure, to obtain a $\Delta U$ map, using a predefined {\em Fitting} function. Then, the encoder ${\mathcal{E}}_{\Delta U}$ that has been learned is used to obtain the reduced $r\Delta U$, which can be input to the learned Strain Model. 
A reduced $r\epsilon_p$ is obtained, which is decoded using ${\mathcal{D}}_{\epsilon_p}$ into a full dimensionality strain map on the original mesh of the structure, that can be interpreted by experts.

FDF facilitates the design of this pipeline in three ways.
First, it allows describing visually the dataflow necessary to learn a function, which is non-standard here as it involves routing an output $\Delta U$ of the first Abaqus Model as the input of the DTP. Second, it enables the easy export of functions from the FDF learning pipeline and their easy reuse at the correct place in the exploitation pipeline. Lastly, implicit typing helps to prevent user mistakes. For example, it can prevent users from mistakenly feeding $\Delta U$ to the Strain Model instead of the expected reduced $r\Delta U$ in case they forget to use ${\mathcal{E}}_{\Delta U}$.

\subsection{DTI of a Magnetic Bearing Instance}

\begin{figure}[b!]
    \centering
    \includegraphics[width=\columnwidth]{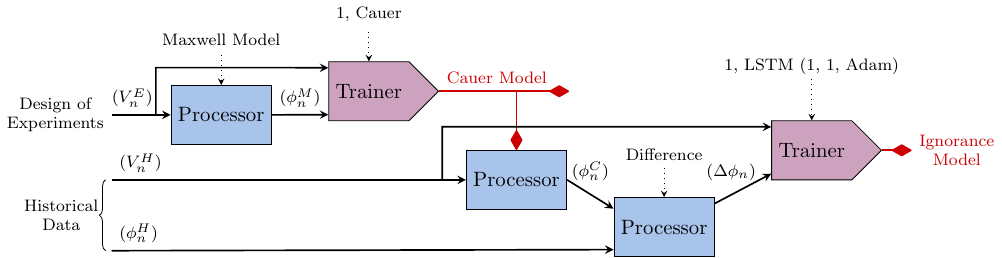}
    \caption{An FDF pipeline to learn a Magnetic Bearing DTI.}
    \label{fig:pipe_skf_offline}
    \Description{The image shows a training pipeline to predict the behavior of a magnetic bearing given an applied voltage. Detailed description in text.}
\end{figure}

Recall that the second motivating example (see Figure~\ref{fig:magnetic-bearing-overview}) aims to predict the magnetic flux given an applied voltage profile, in a particular instance of a bearing~\cite{ghnatiosHybridTwinBased2024}. Figure~\ref{fig:pipe_skf_offline} describes an FDF pipeline to generate the nominal DTP and then the DTI by tuning to the particular instance of a bearing. 

From a design of experiments on various voltage time series $(V_n^E)$ (different amplitude and shape), we obtain, using the (slow) Maxwell's equation (first Processor box), the output sequence $(\phi_n^M)$ of the magnetic flux induced in the system.
A (fast) Cauer model is trained (first Trainer box) from these input-output time series to generalize them accurately. This gives the nominal Cauer model.

To account for the characteristics of a specific magnetic bearing instance (deviations from the nominal model during manufacturing or operation), we introduce an Ignorance Model~\cite{chinestaVirtualDigitalHybrid2020}. This model captures the difference between the nominal Cauer model and the instance's actual behavior observed through historical data collected from it. 
Specifically, we obtain the predicted flux time series $(\phi^C_n)$ from the nominal Cauer model based on some historical voltage $(V_n^H)$. We then compare this predicted flux with the associated historical flux $(\phi_n^H)$ measured on the specific magnetic bearing instance. The last Processor box computes the (time series) difference between the predicted and actual historical behavior. Lastly, we train a Long Short-Term Memory (LSTM) network on these supervised pairs $((V_n^H),(\Delta\phi_n))$, resulting in the Ignorance model.

Note that FDF makes the design easy, in particular for manipulating and reusing the learned functions: the Cauer model is reused to infer flux from data $(V^H_n)$ different from the one it has been learned on $(V_n^E)$. This is unlike the encoder ${\mathcal{E}}_{\Delta U}$ in Figure~\ref{fig:pipe_cetim_offline} that is used on the same dataset that it has been trained with. Hence, not decoupling learning and inference would hinder such generality. Type-checking would also help the designer here. Note that superscripts $^E$ and $^H$ in the annotation are disregarded for the type equality check, and the type propagation will know that both have the same type from annotations.

\balance %

\subsubsection*{Exploitation}
\begin{figure}[t!]
    \centering
    \includegraphics[width=0.6\columnwidth]{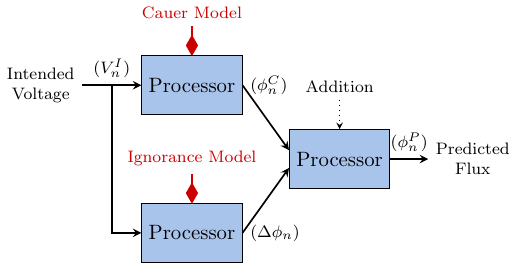}
    \caption{The exploitation pipeline for a Magnetic Bearing.}
    \label{fig:pipe_skf_online}
    \Description{The image shows the final DTI for magnetic bearing using three processor boxes to integrate the Cauer Model and the sensor data.}
\end{figure}

Figure~\ref{fig:pipe_skf_online} details the exploitation pipeline. This pipeline aims to predict the magnetic flux $\phi_n^P$ that would be triggered on the particular instance of the magnetic bearing based on an intended voltage input profile. This prediction allows us to assess whether the bearing will operate as intended with the given voltage profile or if the voltage profile needs to be adjusted.

The pipeline proceeds as follows. The intended voltage time series $(V_n^I)$ is fed into both the learned Cauer and Ignorance models. The Cauer model predicts the nominal flux response $(\phi_n^C)$, while the Ignorance model predicts the discrepancy $(\Delta \phi_n)$ between this particular instance and the nominal Cauer model. These two components are then integrated (by a simple addition $(\phi^C_n + \Delta \phi_n)$) to obtain the predicted DTI flux.

\subsubsection*{Alternate DTI pipeline}
The pipeline in Figure~\ref{fig:pipe_skf_offline} is very accurate when the discrepancy between the instance and the nominal Cauer model directly depends upon the applied voltage. For example, the higher the voltage, the more the instance flux deviates from the nominal Cauer model.
However, in other types of discrepancies, there may be a more complex interplay, e.g., thresholding of instance's flux wrt to the Cauer model. In such cases, employing different operators, such as composition instead of difference, could lead to a more accurate model: the Cauer flux $(\phi_n^C)$ is directly linked to the discrepancy of the flux of the instance in this case.

\begin{figure}[H]
    \centering
    \includegraphics[width=\columnwidth]{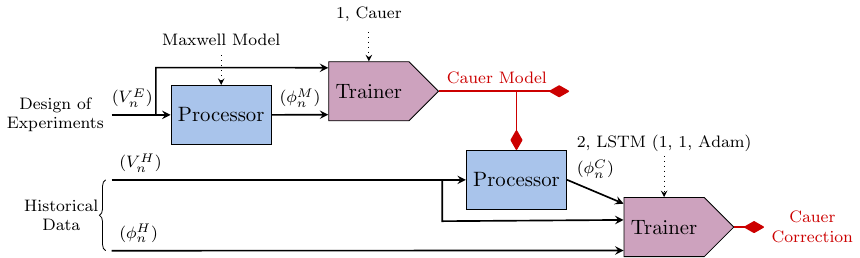}
    \caption{A variant of the Magnetic Bearing DTI pipeline.}
    \label{fig:pipe_skf_offline_var}
    \Description{The image shows a training pipeline to predict the behavior of a magnetic bearing given an applied voltage. Detailed description in text.}
\end{figure}

We finally demonstrate once more the generality and ease of using FDF, by proposing an alternate DTI pipeline to specify a pipeline using composition instead of difference. We illustrate such a variant of the FDF pipeline in Figure~\ref{fig:pipe_skf_offline_var}. 

\begin{figure}[H]
    \centering
    \includegraphics[width=0.6\columnwidth]{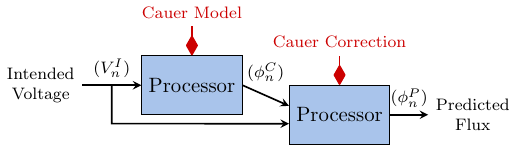}
    \caption{Variant of the exploitation pipeline for a Magnetic Bearing with composition.}
    \label{fig:pipe_skf_online_var}
    \Description{The image shows a training pipeline to predict the behavior of a magnetic bearing given an applied voltage. The Cauer Model receives the voltage and outputs the magnetic flux. Both the voltage and the magnetic flux are provided to the Cauer Correction box to predict the flux in the bearing.}
\end{figure}

The Cauer model is obtained using the same process as before. However, the second Trainer box now has as additional input $(\phi_n^C)$ on top of $(V_n^H)$. This is reflected in the "2" in the first component of its parameter. Similarly, we adapt the exploitation pipeline (see Figure ~\ref{fig:pipe_skf_online_var}): now the Cauer correction is applied {\em after} the Cauer model's prediction, and it takes the flux $(\phi^C_n)$ as input to make its final prediction, thus implementing composition.

\section{Conclusion}\label{sec:conclusion}
This paper introduces the Function+Data Flow (FDF) language, a novel domain-specific language designed to streamline the creation of machine learning pipelines for digital twins.
FDF addresses shortcomings in existing formalisms: 
{\em generality} to twin in many different domains and applications is hindered when the pipeline is rigid and proprietary 
(twin builders from software vendors specialized in CAD/CFD) and
{\em manipulation of functions} is hard when they are not represented explicitly (traditional machine learning workflows).

FDF bridges the first gap by leveraging open machine learning workflows and concepts of dataflow. This flexibility allows FDF to encompass the diverse requirements of various digital twinning domains, as showcased in two motivating examples in electromagnetic and structural engineering. 
To bridge the second gap, FDF extends the classical dataflow by treating functions as first-class citizens. This enables users to manipulate and combine these functions freely, a crucial aspect of digital twinning.
Furthermore, FDF enjoys the maintainability and debuggability of typed languages, through the introduction of user-friendly implicit typing, which removes the burden of explicitly specifying every data type in the pipeline. In essence, FDF integrates the well-established software engineering principles of abstraction into the design of machine learning pipelines for digital twinning.

Future work shall include developing the FDF framework, e.g. through complete libraries, and by extending FDF to support control applications and online learning.

\begin{acks}
We thank our reviewers for their constructive feedback. We thank Amine Ammar and Joel Moutarde for their suggestions and inputs on the motivating examples. 
This research is part of the program DesCartes and is supported by the National Research Foundation, Prime Minister's Office, Singapore under its Campus for Research Excellence and Technological Enterprise (CREATE) program. 
\end{acks}

\bibliographystyle{ACM-Reference-Format}
\bibliography{refs}


\begin{thebibliography}{36}


\ifx \showCODEN    \undefined \def \showCODEN     #1{\unskip}     \fi
\ifx \showDOI      \undefined \def \showDOI       #1{#1}\fi
\ifx \showISBNx    \undefined \def \showISBNx     #1{\unskip}     \fi
\ifx \showISBNxiii \undefined \def \showISBNxiii  #1{\unskip}     \fi
\ifx \showISSN     \undefined \def \showISSN      #1{\unskip}     \fi
\ifx \showLCCN     \undefined \def \showLCCN      #1{\unskip}     \fi
\ifx \shownote     \undefined \def \shownote      #1{#1}          \fi
\ifx \showarticletitle \undefined \def \showarticletitle #1{#1}   \fi
\ifx \showURL      \undefined \def \showURL       {\relax}        \fi
\providecommand\bibfield[2]{#2}
\providecommand\bibinfo[2]{#2}
\providecommand\natexlab[1]{#1}
\providecommand\showeprint[2][]{arXiv:#2}

\bibitem[Alam et~al\mbox{.}(2024)]%
        {alamKedro2024}
\bibfield{author}{\bibinfo{person}{Sajid Alam}, \bibinfo{person}{Nok~Lam Chan}, \bibinfo{person}{Laura Couto}, \bibinfo{person}{Yetunde Dada}, \bibinfo{person}{Ivan Danov}, \bibinfo{person}{Deepyaman Datta}, \bibinfo{person}{Tynan DeBold}, \bibinfo{person}{Jitendra Gundaniya}, \bibinfo{person}{Yolan Honoré-Rougé}, \bibinfo{person}{Stephanie Kaiser}, \bibinfo{person}{Rashida Kanchwala}, \bibinfo{person}{Ankita Katiyar}, \bibinfo{person}{Ravi~Kumar Pilla}, \bibinfo{person}{Huong Nguyen}, \bibinfo{person}{Nero Okwa}, \bibinfo{person}{Juan~Luis Cano~Rodríguez}, \bibinfo{person}{Joel Schwarzmann}, \bibinfo{person}{Dmitry Sorokin}, \bibinfo{person}{Merel Theisen}, \bibinfo{person}{Marcin Zabłocki}, {and} \bibinfo{person}{Simon Brugman}.} \bibinfo{year}{2024}\natexlab{}.
\newblock \bibinfo{booktitle}{\emph{{Kedro}}}.
\newblock Kedro.
\newblock
\urldef\tempurl%
\url{https://github.com/kedro-org/kedro}
\showURL{%
\tempurl}


\bibitem[American Society of Mechanical Engineers(2023)]%
        {johnsonComplexitiesCapturingLarge2023}
American Society of Mechanical Engineers \bibinfo{year}{2023}\natexlab{}.
\newblock \bibinfo{booktitle}{\emph{{Complexities of Capturing Large Plastic Deformations Using Digital Image Correlation: A Test Case on Full-Scale Pipe Specimens}}}. \bibinfo{series}{International Conference on Offshore Mechanics and Arctic Engineering}, Vol.~\bibinfo{volume}{Volume 3: Materials Technology; Pipelines, Risers, and Subsea Systems}. American Society of Mechanical Engineers.
\newblock
\urldef\tempurl%
\url{https://doi.org/10.1115/OMAE2023-102308}
\showDOI{\tempurl}
\showeprint{https://asmedigitalcollection.asme.org/OMAE/proceedings-pdf/OMAE2023/86854/V003T04A033/7040915/v003t04a033-omae2023-102308.pdf}


\bibitem[Ansys(2024)]%
        {AnsysTwinBuilder}
Ansys \bibinfo{year}{2024}\natexlab{}.
\newblock \bibinfo{booktitle}{\emph{Ansys {{Twin Builder}} | {{Create}} and {{Deploy Digital Twin Models}}}}.
\newblock Ansys.
\newblock
\urldef\tempurl%
\url{https://www.ansys.com/products/digital-twin/ansys-twin-builder}
\showURL{%
\tempurl}


\bibitem[Apache Airflow(2024)]%
        {ApacheAirflow}
Apache Airflow \bibinfo{year}{2024}\natexlab{}.
\newblock \bibinfo{booktitle}{\emph{Apache {{Airflow}}}}.
\newblock Apache Airflow.
\newblock
\urldef\tempurl%
\url{https://airflow.apache.org/}
\showURL{%
\tempurl}


\bibitem[Bogner and Merkel(2022)]%
        {bognerTypeNotType2022}
\bibfield{author}{\bibinfo{person}{Justus Bogner} {and} \bibinfo{person}{Manuel Merkel}.} \bibinfo{year}{2022}\natexlab{}.
\newblock \showarticletitle{To Type or Not to Type? A Systematic Comparison of the Software Quality of {{JavaScript}} and Typescript Applications on {{GitHub}}}. In \bibinfo{booktitle}{\emph{Proceedings of the 19th {{International Conference}} on {{Mining Software Repositories}}}} (2022-10-17) \emph{(\bibinfo{series}{{{MSR}} '22})}. \bibinfo{publisher}{Association for Computing Machinery}, \bibinfo{address}{New York, NY, USA}, \bibinfo{pages}{658--669}.
\newblock
\showISBNx{978-1-4503-9303-4}
\urldef\tempurl%
\url{https://doi.org/10.1145/3524842.3528454}
\showDOI{\tempurl}


\bibitem[Chabod(2022)]%
        {chabodDigitalTwinFatigue2022}
\bibfield{author}{\bibinfo{person}{Amaury Chabod}.} \bibinfo{year}{2022}\natexlab{}.
\newblock \showarticletitle{Digital {{Twin}} for {{Fatigue Analysis}}}.
\newblock \bibinfo{journal}{\emph{Procedia Structural Integrity}}  \bibinfo{volume}{38} (\bibinfo{date}{Jan.} \bibinfo{year}{2022}), \bibinfo{pages}{382--392}.
\newblock
\showISSN{2452-3216}
\urldef\tempurl%
\url{https://doi.org/10.1016/j.prostr.2022.03.039}
\showDOI{\tempurl}


\bibitem[Chen et~al\mbox{.}(2020)]%
        {chenDevelopmentsMLflowSystem2020}
\bibfield{author}{\bibinfo{person}{Andrew Chen}, \bibinfo{person}{Andy Chow}, \bibinfo{person}{Aaron Davidson}, \bibinfo{person}{Arjun DCunha}, \bibinfo{person}{Ali Ghodsi}, \bibinfo{person}{Sue~Ann Hong}, \bibinfo{person}{Andy Konwinski}, \bibinfo{person}{Clemens Mewald}, \bibinfo{person}{Siddharth Murching}, \bibinfo{person}{Tomas Nykodym}, \bibinfo{person}{Paul Ogilvie}, \bibinfo{person}{Mani Parkhe}, \bibinfo{person}{Avesh Singh}, \bibinfo{person}{Fen Xie}, \bibinfo{person}{Matei Zaharia}, \bibinfo{person}{Richard Zang}, \bibinfo{person}{Juntai Zheng}, {and} \bibinfo{person}{Corey Zumar}.} \bibinfo{year}{2020}\natexlab{}.
\newblock \showarticletitle{Developments in MLflow: A System to Accelerate the Machine Learning Lifecycle}. In \bibinfo{booktitle}{\emph{Proceedings of the Fourth International Workshop on Data Management for End-to-End Machine Learning}} (Portland, OR, USA) \emph{(\bibinfo{series}{DEEM'20})}. \bibinfo{publisher}{Association for Computing Machinery}, \bibinfo{address}{New York, NY, USA}, Article \bibinfo{articleno}{5}, \bibinfo{numpages}{4}~pages.
\newblock
\showISBNx{9781450380232}
\urldef\tempurl%
\url{https://doi.org/10.1145/3399579.3399867}
\showDOI{\tempurl}


\bibitem[Chinesta et~al\mbox{.}(2020)]%
        {chinestaVirtualDigitalHybrid2020}
\bibfield{author}{\bibinfo{person}{Francisco Chinesta}, \bibinfo{person}{Elias Cueto}, \bibinfo{person}{Emmanuelle Abisset-Chavanne}, \bibinfo{person}{Jean~Louis Duval}, {and} \bibinfo{person}{Fouad~El Khaldi}.} \bibinfo{year}{2020}\natexlab{}.
\newblock \showarticletitle{Virtual, {{Digital}} and {{Hybrid Twins}}: {{A New Paradigm}} in {{Data-Based Engineering}} and {{Engineered Data}}}.
\newblock \bibinfo{journal}{\emph{Arch Computat Methods Eng}} \bibinfo{volume}{27}, \bibinfo{number}{1} (\bibinfo{date}{Jan.} \bibinfo{year}{2020}), \bibinfo{pages}{105--134}.
\newblock
\showISSN{1134-3060, 1886-1784}
\urldef\tempurl%
\url{https://doi.org/10.1007/s11831-018-9301-4}
\showDOI{\tempurl}


\bibitem[Chinesta et~al\mbox{.}(2011)]%
        {chinestaShortReviewModel2011}
\bibfield{author}{\bibinfo{person}{Francisco Chinesta}, \bibinfo{person}{Pierre Ladeveze}, {and} \bibinfo{person}{Elías Cueto}.} \bibinfo{year}{2011}\natexlab{}.
\newblock \showarticletitle{A {{Short Review}} on {{Model Order Reduction Based}} on {{Proper Generalized Decomposition}}}.
\newblock \bibinfo{journal}{\emph{Arch Computat Methods Eng}} \bibinfo{volume}{18}, \bibinfo{number}{4} (\bibinfo{date}{Nov.} \bibinfo{year}{2011}), \bibinfo{pages}{395--404}.
\newblock
\showISSN{1134-3060, 1886-1784}
\urldef\tempurl%
\url{https://doi.org/10.1007/s11831-011-9064-7}
\showDOI{\tempurl}


\bibitem[Danilczyk et~al\mbox{.}(2019)]%
        {danilczykANGELIntelligentDigital2019}
\bibfield{author}{\bibinfo{person}{William Danilczyk}, \bibinfo{person}{Yan Sun}, {and} \bibinfo{person}{Haibo He}.} \bibinfo{year}{2019}\natexlab{}.
\newblock \showarticletitle{{{ANGEL}}: {{An Intelligent Digital Twin Framework}} for {{Microgrid Security}}}. In \bibinfo{booktitle}{\emph{2019 {{North American Power Symposium}} ({{NAPS}})}}. \bibinfo{publisher}{IEEE}, \bibinfo{address}{Wichita, KS, USA}, \bibinfo{pages}{1--6}.
\newblock
\showISBNx{978-1-72810-407-2}
\urldef\tempurl%
\url{https://doi.org/10.1109/NAPS46351.2019.9000371}
\showDOI{\tempurl}


\bibitem[De~Meo and Homer(2022)]%
        {demeoDomainSpecificVisualLanguage2022}
\bibfield{author}{\bibinfo{person}{Alexis De~Meo} {and} \bibinfo{person}{Michael Homer}.} \bibinfo{year}{2022}\natexlab{}.
\newblock \showarticletitle{Domain-Specific Visual Language for Data Engineering Quality}. In \bibinfo{booktitle}{\emph{Proceedings of the 1st ACM SIGPLAN International Workshop on Programming Abstractions and Interactive Notations, Tools, and Environments}} (Auckland, New Zealand) \emph{(\bibinfo{series}{PAINT 2022})}. \bibinfo{publisher}{Association for Computing Machinery}, \bibinfo{address}{New York, NY, USA}, \bibinfo{pages}{48–56}.
\newblock
\showISBNx{9781450399104}
\urldef\tempurl%
\url{https://doi.org/10.1145/3563836.3568727}
\showDOI{\tempurl}


\bibitem[Ferrise et~al\mbox{.}(2013)]%
        {ferriseInteractiveVirtualPrototypes2013}
\bibfield{author}{\bibinfo{person}{Francesco Ferrise}, \bibinfo{person}{Monica Bordegoni}, {and} \bibinfo{person}{Umberto Cugini}.} \bibinfo{year}{2013}\natexlab{}.
\newblock \showarticletitle{Interactive {{Virtual Prototypes}} for {{Testing}} the {{Interaction}} with New {{Products}}}.
\newblock \bibinfo{journal}{\emph{Computer-Aided Design and Applications}} \bibinfo{volume}{10}, \bibinfo{number}{3} (\bibinfo{date}{Jan.} \bibinfo{year}{2013}), \bibinfo{pages}{515--525}.
\newblock
\showISSN{null}
\urldef\tempurl%
\url{https://doi.org/10.3722/cadaps.2013.515-525}
\showDOI{\tempurl}


\bibitem[Fortune Business Insights(2024)]%
        {DigitalTwinMarket}
Fortune Business Insights \bibinfo{year}{2024}\natexlab{}.
\newblock \bibinfo{booktitle}{\emph{Digital {{Twin Market Size}}, {{Share}} | {{Growth Analysis Report}} [2032]}}.
\newblock Fortune Business Insights.
\newblock
\urldef\tempurl%
\url{https://www.fortunebusinessinsights.com/digital-twin-market-106246}
\showURL{%
\tempurl}


\bibitem[Fukunaga et~al\mbox{.}(1993)]%
        {fukunagaFunctionsObjectsData1993}
\bibfield{author}{\bibinfo{person}{Alex Fukunaga}, \bibinfo{person}{Wolfgang Pree}, {and} \bibinfo{person}{Takayuki~Dan Kimura}.} \bibinfo{year}{1993}\natexlab{}.
\newblock \showarticletitle{Functions as Objects in a Data Flow Based Visual Language}. In \bibinfo{booktitle}{\emph{Proceedings of the 1993 {{ACM}} Conference on {{Computer}} Science - {{CSC}} '93}}. \bibinfo{publisher}{ACM Press}, \bibinfo{address}{Indianapolis, Indiana, USA}, \bibinfo{pages}{215--220}.
\newblock
\showISBNx{978-0-89791-558-8}
\urldef\tempurl%
\url{https://doi.org/10.1145/170791.170832}
\showDOI{\tempurl}


\bibitem[Gartner(2022)]%
        {EmergingTechnologiesRevenue}
Gartner \bibinfo{year}{2022}\natexlab{}.
\newblock \bibinfo{booktitle}{\emph{Emerging {{Technologies}}: {{Revenue Opportunity Projection}} of {{Digital Twins}}}}.
\newblock Gartner.
\newblock
\urldef\tempurl%
\url{https://www.gartner.com/en/documents/4011590}
\showURL{%
\tempurl}


\bibitem[Ghnatios et~al\mbox{.}(2024)]%
        {ghnatiosHybridTwinBased2024}
\bibfield{author}{\bibinfo{person}{Chady Ghnatios}, \bibinfo{person}{Sebastian Rodriguez}, \bibinfo{person}{Jerome Tomezyk}, \bibinfo{person}{Yves Dupuis}, \bibinfo{person}{Joel Mouterde}, \bibinfo{person}{Joaquim Da~Silva}, {and} \bibinfo{person}{Francisco Chinesta}.} \bibinfo{year}{2024}\natexlab{}.
\newblock \showarticletitle{A Hybrid Twin Based on Machine Learning Enhanced Reduced Order Model for Real-Time Simulation of Magnetic Bearings}.
\newblock \bibinfo{journal}{\emph{Adv. Model. and Simul. in Eng. Sci.}} \bibinfo{volume}{11}, \bibinfo{number}{1} (\bibinfo{year}{2024}), \bibinfo{pages}{3}.
\newblock
\showISSN{2213-7467}
\urldef\tempurl%
\url{https://doi.org/10.1186/s40323-024-00258-2}
\showDOI{\tempurl}


\bibitem[Grieves and Vickers(2017)]%
        {grievesDigitalTwinMitigating2017}
\bibfield{author}{\bibinfo{person}{Michael Grieves} {and} \bibinfo{person}{John Vickers}.} \bibinfo{year}{2017}\natexlab{}.
\newblock \bibinfo{booktitle}{\emph{Digital Twin: Mitigating Unpredictable, Undesirable Emergent Behavior in Complex Systems}}.
\newblock \bibinfo{publisher}{Springer International Publishing}, \bibinfo{address}{Cham}, \bibinfo{pages}{85--113}.
\newblock
\showISBNx{978-3-319-38756-7}
\urldef\tempurl%
\url{https://doi.org/10.1007/978-3-319-38756-7_4}
\showDOI{\tempurl}


\bibitem[Hartmann et~al\mbox{.}(2018)]%
        {hartmannModelOrderReduction2018}
\bibfield{author}{\bibinfo{person}{Dirk Hartmann}, \bibinfo{person}{Matthias Herz}, {and} \bibinfo{person}{Utz Wever}.} \bibinfo{year}{2018}\natexlab{}.
\newblock \showarticletitle{Model {{Order Reduction}} a {{Key Technology}} for {{Digital Twins}}}.
\newblock In \bibinfo{booktitle}{\emph{Reduced-{{Order Modeling}} ({{ROM}}) for {{Simulation}} and {{Optimization}}: {{Powerful Algorithms}} as {{Key Enablers}} for {{Scientific Computing}}}}, \bibfield{editor}{\bibinfo{person}{Winfried Keiper}, \bibinfo{person}{Anja Milde}, {and} \bibinfo{person}{Stefan Volkwein}} (Eds.). \bibinfo{publisher}{Springer International Publishing}, \bibinfo{address}{Cham}, \bibinfo{pages}{167--179}.
\newblock
\showISBNx{978-3-319-75319-5}
\urldef\tempurl%
\url{https://doi.org/10.1007/978-3-319-75319-5_8}
\showDOI{\tempurl}


\bibitem[Jafari et~al\mbox{.}(2023)]%
        {jafariReviewDigitalTwin2023}
\bibfield{author}{\bibinfo{person}{Mina Jafari}, \bibinfo{person}{Abdollah Kavousi-Fard}, \bibinfo{person}{Tao Chen}, {and} \bibinfo{person}{Mazaher Karimi}.} \bibinfo{year}{2023}\natexlab{}.
\newblock \showarticletitle{A {{Review}} on {{Digital Twin Technology}} in {{Smart Grid}}, {{Transportation System}} and {{Smart City}}: {{Challenges}} and {{Future}}}.
\newblock \bibinfo{journal}{\emph{IEEE Access}}  \bibinfo{volume}{11} (\bibinfo{year}{2023}), \bibinfo{pages}{17471--17484}.
\newblock
\showISSN{2169-3536}
\urldef\tempurl%
\url{https://doi.org/10.1109/ACCESS.2023.3241588}
\showDOI{\tempurl}


\bibitem[Johnston et~al\mbox{.}(2004)]%
        {johnstonAdvancesDataflowProgramming2004}
\bibfield{author}{\bibinfo{person}{Wesley~M. Johnston}, \bibinfo{person}{J.~R.~Paul Hanna}, {and} \bibinfo{person}{Richard~J. Millar}.} \bibinfo{year}{2004}\natexlab{}.
\newblock \showarticletitle{Advances in Dataflow Programming Languages}.
\newblock \bibinfo{journal}{\emph{ACM Comput. Surv.}} \bibinfo{volume}{36}, \bibinfo{number}{1} (\bibinfo{date}{March} \bibinfo{year}{2004}), \bibinfo{pages}{1--34}.
\newblock
\showISSN{0360-0300}
\urldef\tempurl%
\url{https://doi.org/10.1145/1013208.1013209}
\showDOI{\tempurl}


\bibitem[Kritzinger et~al\mbox{.}(2018)]%
        {kritzingerDigitalTwinManufacturing2018a}
\bibfield{author}{\bibinfo{person}{Werner Kritzinger}, \bibinfo{person}{Matthias Karner}, \bibinfo{person}{Georg Traar}, \bibinfo{person}{Jan Henjes}, {and} \bibinfo{person}{Wilfried Sihn}.} \bibinfo{year}{2018}\natexlab{}.
\newblock \showarticletitle{Digital {{Twin}} in Manufacturing: {{A}} Categorical Literature Review and Classification}.
\newblock \bibinfo{journal}{\emph{IFAC-PapersOnLine}} \bibinfo{volume}{51}, \bibinfo{number}{11} (\bibinfo{year}{2018}), \bibinfo{pages}{1016--1022}.
\newblock
\showISSN{24058963}
\urldef\tempurl%
\url{https://doi.org/10.1016/j.ifacol.2018.08.474}
\showDOI{\tempurl}


\bibitem[Kubeflow(2024)]%
        {Kubeflow}
Kubeflow \bibinfo{year}{2024}\natexlab{}.
\newblock \bibinfo{booktitle}{\emph{Kubeflow}}.
\newblock Kubeflow.
\newblock
\urldef\tempurl%
\url{https://www.kubeflow.org/}
\showURL{%
\tempurl}


\bibitem[Lwakatare et~al\mbox{.}(2020)]%
        {lwakatareDataScienceDriven2020}
\bibfield{author}{\bibinfo{person}{Lucy~Ellen Lwakatare}, \bibinfo{person}{Ivica Crnkovic}, \bibinfo{person}{Ellinor Rånge}, {and} \bibinfo{person}{Jan Bosch}.} \bibinfo{year}{2020}\natexlab{}.
\newblock \showarticletitle{From a {{Data Science Driven Process}} to a {{Continuous Delivery Process}} for {{Machine Learning Systems}}}.
\newblock In \bibinfo{booktitle}{\emph{Product-{{Focused Software Process Improvement}}}}, \bibfield{editor}{\bibinfo{person}{Maurizio Morisio}, \bibinfo{person}{Marco Torchiano}, {and} \bibinfo{person}{Andreas Jedlitschka}} (Eds.). Vol.~\bibinfo{volume}{12562}. \bibinfo{publisher}{Springer International Publishing}, \bibinfo{address}{Cham}, \bibinfo{pages}{185--201}.
\newblock
\showISBNx{978-3-030-64148-1}
\urldef\tempurl%
\url{https://doi.org/10.1007/978-3-030-64148-1_12}
\showDOI{\tempurl}


\bibitem[Moya et~al\mbox{.}(7 15)]%
        {moyaDigitalTwinsThat2022}
\bibfield{author}{\bibinfo{person}{Beatriz Moya}, \bibinfo{person}{Alberto Badías}, \bibinfo{person}{Icíar Alfaro}, \bibinfo{person}{Francisco Chinesta}, {and} \bibinfo{person}{Elías Cueto}.} \bibinfo{year}{2022-07-15}\natexlab{}.
\newblock \showarticletitle{Digital Twins That Learn and Correct Themselves}.
\newblock \bibinfo{journal}{\emph{Numerical Meth Engineering}} \bibinfo{volume}{123}, \bibinfo{number}{13} (\bibinfo{year}{2022-07-15}), \bibinfo{pages}{3034--3044}.
\newblock
\showISSN{0029-5981, 1097-0207}
\urldef\tempurl%
\url{https://doi.org/10.1002/nme.6535}
\showDOI{\tempurl}


\bibitem[Ore et~al\mbox{.}(2018)]%
        {oreAssessingTypeAnnotation2018}
\bibfield{author}{\bibinfo{person}{John-Paul Ore}, \bibinfo{person}{Sebastian Elbaum}, \bibinfo{person}{Carrick Detweiler}, {and} \bibinfo{person}{Lambros Karkazis}.} \bibinfo{year}{2018}\natexlab{}.
\newblock \showarticletitle{Assessing the type annotation burden}. In \bibinfo{booktitle}{\emph{Proceedings of the 33rd ACM/IEEE International Conference on Automated Software Engineering}} (Montpellier, France) \emph{(\bibinfo{series}{ASE '18})}. \bibinfo{publisher}{Association for Computing Machinery}, \bibinfo{address}{New York, NY, USA}, \bibinfo{pages}{190–201}.
\newblock
\showISBNx{9781450359375}
\urldef\tempurl%
\url{https://doi.org/10.1145/3238147.3238173}
\showDOI{\tempurl}


\bibitem[Ray et~al\mbox{.}(2014)]%
        {rayLargeScaleStudy2014}
\bibfield{author}{\bibinfo{person}{Baishakhi Ray}, \bibinfo{person}{Daryl Posnett}, \bibinfo{person}{Vladimir Filkov}, {and} \bibinfo{person}{Premkumar Devanbu}.} \bibinfo{year}{2014}\natexlab{}.
\newblock \showarticletitle{A large scale study of programming languages and code quality in github}. In \bibinfo{booktitle}{\emph{Proceedings of the 22nd ACM SIGSOFT International Symposium on Foundations of Software Engineering}} (Hong Kong, China) \emph{(\bibinfo{series}{FSE 2014})}. \bibinfo{publisher}{Association for Computing Machinery}, \bibinfo{address}{New York, NY, USA}, \bibinfo{pages}{155–165}.
\newblock
\showISBNx{9781450330565}
\urldef\tempurl%
\url{https://doi.org/10.1145/2635868.2635922}
\showDOI{\tempurl}


\bibitem[Sancarlos et~al\mbox{.}(2021)]%
        {sancarlosLearningStableReducedorder2021}
\bibfield{author}{\bibinfo{person}{Abel Sancarlos}, \bibinfo{person}{Morgan Cameron}, \bibinfo{person}{Jean-Marc Le~Peuvedic}, \bibinfo{person}{Juliette Groulier}, \bibinfo{person}{Jean-Louis Duval}, \bibinfo{person}{Elias Cueto}, {and} \bibinfo{person}{Francisco Chinesta}.} \bibinfo{year}{2021}\natexlab{}.
\newblock \showarticletitle{Learning Stable Reduced-Order Models for Hybrid Twins}.
\newblock \bibinfo{journal}{\emph{Data-Centric Eng.}}  \bibinfo{volume}{2} (\bibinfo{year}{2021}), \bibinfo{pages}{e10}.
\newblock
\showISSN{2632-6736}
\urldef\tempurl%
\url{https://doi.org/10.1017/dce.2021.16}
\showDOI{\tempurl}


\bibitem[Simcenter(2024)]%
        {SimcenterSystemsSimulation}
Siemens Digital Industries Software \bibinfo{year}{2024}\natexlab{}.
\newblock \bibinfo{booktitle}{\emph{Simcenter Systems Simulation}}.
\newblock Siemens Digital Industries Software.
\newblock
\urldef\tempurl%
\url{https://plm.sw.siemens.com/en-US/simcenter/systems-simulation/}
\showURL{%
\tempurl}


\bibitem[Siva~Srinivas et~al\mbox{.}(2018)]%
        {sivasrinivasApplicationActiveMagnetic2018}
\bibfield{author}{\bibinfo{person}{R. Siva~Srinivas}, \bibinfo{person}{R. Tiwari}, {and} \bibinfo{person}{Ch. Kannababu}.} \bibinfo{year}{2018}\natexlab{}.
\newblock \showarticletitle{Application of Active Magnetic Bearings in Flexible Rotordynamic Systems – {{A}} State-of-the-Art Review}.
\newblock \bibinfo{journal}{\emph{Mechanical Systems and Signal Processing}}  \bibinfo{volume}{106} (\bibinfo{date}{June} \bibinfo{year}{2018}), \bibinfo{pages}{537--572}.
\newblock
\showISSN{08883270}
\urldef\tempurl%
\url{https://doi.org/10.1016/j.ymssp.2018.01.010}
\showDOI{\tempurl}


\bibitem[Sivarajah et~al\mbox{.}(2022)]%
        {sivarajahTierkreisDataflowFramework2022}
\bibfield{author}{\bibinfo{person}{Seyon Sivarajah}, \bibinfo{person}{Lukas Heidemann}, \bibinfo{person}{Alan Lawrence}, {and} \bibinfo{person}{Ross Duncan}.} \bibinfo{year}{2022}\natexlab{}.
\newblock \showarticletitle{Tierkreis: A {{Dataflow Framework}} for {{Hybrid Quantum-Classical Computing}}}. In \bibinfo{booktitle}{\emph{2022 {{IEEE}}/{{ACM Third International Workshop}} on {{Quantum Computing Software}} ({{QCS}})}}. \bibinfo{publisher}{IEEE}, \bibinfo{address}{Dallas, TX, USA}, \bibinfo{pages}{12--21}.
\newblock
\urldef\tempurl%
\url{https://doi.org/10.1109/QCS56647.2022.00007}
\showDOI{\tempurl}


\bibitem[Tehrani et~al\mbox{.}(2020)]%
        {tehraniPipeProfilingUsing2020}
\bibfield{author}{\bibinfo{person}{Amin~Darabnoush Tehrani}, \bibinfo{person}{Zahra Kohankar~Kouchesfehani}, {and} \bibinfo{person}{Mohammad Najafi}.} \bibinfo{year}{2020}\natexlab{}.
\newblock \showarticletitle{Pipe profiling using digital image correlation}.
\newblock In \bibinfo{booktitle}{\emph{Pipelines 2020}}. \bibinfo{publisher}{American Society of Civil Engineers Reston, VA}, \bibinfo{address}{San Antonio, Texas, USA}, \bibinfo{pages}{36--45}.
\newblock


\bibitem[Tuegel et~al\mbox{.}(2011)]%
        {tuegelReengineeringAircraftStructural2011}
\bibfield{author}{\bibinfo{person}{Eric~J. Tuegel}, \bibinfo{person}{Anthony~R. Ingraffea}, \bibinfo{person}{Thomas~G. Eason}, {and} \bibinfo{person}{S.~Michael Spottswood}.} \bibinfo{year}{2011}\natexlab{}.
\newblock \showarticletitle{Reengineering {{Aircraft Structural Life Prediction Using}} a {{Digital Twin}}}.
\newblock \bibinfo{journal}{\emph{International Journal of Aerospace Engineering}}  \bibinfo{volume}{2011} (\bibinfo{date}{Oct.} \bibinfo{year}{2011}), \bibinfo{pages}{e154798}.
\newblock
\showISSN{1687-5966}
\urldef\tempurl%
\url{https://doi.org/10.1155/2011/154798}
\showDOI{\tempurl}


\bibitem[Utzig et~al\mbox{.}(2019)]%
        {utzigAugmentedRealityRemote2019}
\bibfield{author}{\bibinfo{person}{Sebastian Utzig}, \bibinfo{person}{Robert Kaps}, \bibinfo{person}{Syed~Muhammad Azeem}, {and} \bibinfo{person}{Andreas Gerndt}.} \bibinfo{year}{2019}\natexlab{}.
\newblock \showarticletitle{Augmented {{Reality}} for {{Remote Collaboration}} in {{Aircraft Maintenance Tasks}}}. In \bibinfo{booktitle}{\emph{2019 {{IEEE Aerospace Conference}}}}. \bibinfo{publisher}{IEEE}, \bibinfo{address}{Big Sky, MT, USA}, \bibinfo{pages}{1--10}.
\newblock
\showISBNx{978-1-5386-6854-2}
\urldef\tempurl%
\url{https://doi.org/10.1109/AERO.2019.8742228}
\showDOI{\tempurl}


\bibitem[Wang et~al\mbox{.}(2023)]%
        {wangShortTermWindSpeed2023}
\bibfield{author}{\bibinfo{person}{Zhongju Wang}, \bibinfo{person}{Long Wang}, \bibinfo{person}{M Revanesh}, \bibinfo{person}{Chao Huang}, {and} \bibinfo{person}{Xiong Luo}.} \bibinfo{year}{2023}\natexlab{}.
\newblock \showarticletitle{Short-{{Term Wind Speed}} and {{Power Forecasting}} for {{Smart City Power Grid With}} a {{Hybrid Machine Learning Framework}}}.
\newblock \bibinfo{journal}{\emph{IEEE Internet Things J.}} \bibinfo{volume}{10}, \bibinfo{number}{21} (\bibinfo{date}{Nov.} \bibinfo{year}{2023}), \bibinfo{pages}{18754--18765}.
\newblock
\showISSN{2327-4662, 2372-2541}
\urldef\tempurl%
\url{https://doi.org/10.1109/JIOT.2023.3286568}
\showDOI{\tempurl}


\bibitem[Wolf et~al\mbox{.}(2020)]%
        {wolfHuggingFaceTransformersStateoftheart2020}
\bibfield{author}{\bibinfo{person}{Thomas Wolf}, \bibinfo{person}{Lysandre Debut}, \bibinfo{person}{Victor Sanh}, \bibinfo{person}{Julien Chaumond}, \bibinfo{person}{Clement Delangue}, \bibinfo{person}{Anthony Moi}, \bibinfo{person}{Pierric Cistac}, \bibinfo{person}{Tim Rault}, \bibinfo{person}{Rémi Louf}, \bibinfo{person}{Morgan Funtowicz}, \bibinfo{person}{Joe Davison}, \bibinfo{person}{Sam Shleifer}, \bibinfo{person}{Patrick von Platen}, \bibinfo{person}{Clara Ma}, \bibinfo{person}{Yacine Jernite}, \bibinfo{person}{Julien Plu}, \bibinfo{person}{Canwen Xu}, \bibinfo{person}{Teven~Le Scao}, \bibinfo{person}{Sylvain Gugger}, \bibinfo{person}{Mariama Drame}, \bibinfo{person}{Quentin Lhoest}, {and} \bibinfo{person}{Alexander~M. Rush}.} \bibinfo{year}{2020}\natexlab{}.
\newblock \bibinfo{title}{HuggingFace's Transformers: State-of-the-art Natural Language Processing}.
\newblock
\newblock
\showeprint[arxiv]{1910.03771}~[cs.CL]


\bibitem[Xiong and Wang(2022)]%
        {xiongDigitalTwinApplications2022}
\bibfield{author}{\bibinfo{person}{Minglan Xiong} {and} \bibinfo{person}{Huawei Wang}.} \bibinfo{year}{2022}\natexlab{}.
\newblock \showarticletitle{Digital Twin Applications in Aviation Industry: {{A}} Review}.
\newblock \bibinfo{journal}{\emph{Int J Adv Manuf Technol}} \bibinfo{volume}{121}, \bibinfo{number}{9} (\bibinfo{year}{2022}), \bibinfo{pages}{5677--5692}.
\newblock
Issue 9.
\showISSN{1433-3015}
\urldef\tempurl%
\url{https://doi.org/10.1007/s00170-022-09717-9}
\showDOI{\tempurl}


\end{thebibliography}

\end{document}